\documentstyle[12pt]{article}

\newcommand{\be}{\begin{equation}}
\newcommand{\ee}{\end{equation}}
\newcommand{\bea}{\begin{eqnarray}}
\newcommand{\eea}{\end{eqnarray}}
\newcommand{\beasn}{\begin{sneqnarray}}
\newcommand{\eeasn}{\end{sneqnarray}}
\newcommand{\bref}[1]{(\ref{#1})}
\newcommand{\eps}{\epsilon}
\newcommand{\veps}{\varepsilon}

\newcommand{\der}[2]{\frac{\partial #1}{\partial #2}}
\newcommand{\lder}[2]{\frac{\partial_l#1}{\partial #2}}
\newcommand{\rder}[2]{\frac{\partial_r#1}{\partial #2}}
\newcommand{\gh}[1]{{\cal #1}}

\def\fnote#1#2{\begingroup\def\thefootnote{#1}\footnote{#2}\addtocounter
{footnote}{-1}\endgroup}

\makeatletter

\def\mathoperat{\@ifnextchar [{\@mathoperat}{\@mathoperat[rm]}}
   \def\@mathoperat[#1]#2#3{\def#2{\mathop{\@nameuse{#1} #3{}}\nolimits}}

\def\dif{{\rm d}}
\def\deriv{\@ifnextchar[{\@deriv}{\@deriv[]}}
   \def\@deriv[#1]#2#3{\mathchoice%
{{\dif^{#1}#2\over\dif{#3}^{#1}}}{{\dif^{#1}#2/\dif{#3}^{#1}}}%
{{\dif^{#1}#2\over\dif{#3}^{#1}}}{{\dif^{#1}#2/\dif{#3}^{#1}}}}
\def\derpar#1#2{\mathchoice%
{{\partial#1\over\partial#2}}{{\partial#1/\partial#2}}%
{{\partial#1\over\partial#2}}{{\partial#1/\partial#2}}}

\def\secteqno{\@addtoreset{equation}{section}%
\def\theequation{\thesection.\arabic{equation}}}
\def\endsecteqno{\def\theequation{\@ifundefined{chapter}%
{\arabic{equation}}{\thechapter.\arabic{equation}}}}
\newcounter{subequation}
\def\thesubequation{\alph{subequation}}
\def\sneqnarray{\stepcounter{equation}\let\@currentlabel=\theequation
\setcounter{subequation}{1}
\def\@eqnnum{{\rm (\theequation.\thesubequation)}}
\global\@eqcnt\z@\tabskip\@centering\let\\=\@eqncr\let\@@eqncr=\@@sneqncr
$$\halign to \displaywidth\bgroup\@eqnsel\hskip\@centering
 $\displaystyle\tabskip\z@{##}$&\global\@eqcnt\@ne
 \hskip 2\arraycolsep \hfil${##}$\hfil
 &\global\@eqcnt\tw@ \hskip 2\arraycolsep $\displaystyle\tabskip\z@{##}$\hfil
  \tabskip\@centering&\llap{##}\tabskip\z@\cr}
\def\endsneqnarray{\@@sneqncr\egroup $$\global\@ignoretrue}
\def\@@sneqncr{\let\@tempa\relax
   \ifcase\@eqcnt \def\@tempa{& & &}\or \def\@tempa{& &}
   \else \def\@tempa{&}\fi
     \@tempa \if@eqnsw\@eqnnum\stepcounter{subequation}\fi
     \global\@eqnswtrue\global\@eqcnt\z@\cr}
\def\nobiblabels{\def\@lbibitem[##1]##2{\@bibitem{##2}}}
\def\tr{\mathop{\rm tr}\nolimits}

\makeatother

\secteqno
\begin{document}

\hfill UUB-ECM-PF 93/5

\hfill UTTG-02-93

\hfill February 1993

\begin{center}

{\large{\bf BV analysis for covariant and non-covariant actions}}

\vspace{24pt}

C.\,Ord\'o\~nez\fnote{*}{Bitnet address:
                  Cordonez\,@\,UTAPHY, World Laboratory Fellow}

\vspace{10pt}
{\it Theory Group\\ Department of Physics \\ University of Texas
\\ Austin, Texas 78712}

\vspace{14pt}

J. Par\'{\i}s\fnote{ \dagger }{ Bitnet address: Paris@EBUBECM1 },
   J. M. Pons\fnote{\ddagger}{ Bitnet address: Pons@EBUBECM1 }
        and   R. Toldr\`a

\vspace{10pt}
       {\it Departament d'Estructura i Constituents de la
                   Mat\`eria\\
      Facultat de F\'{\i}sica\\
      Universitat de Barcelona\\
       Diagonal 647\\
        E-08028 Barcelona\\
        Catalonia}

\normalsize
 \end{center}

\pagestyle{myheadings}

\markboth{}{\it C.\,Ordo\~nez, J.\,Par\'{\i}s, J.\,M.\,Pons and
R. Toldr\`a  ~~{\rm BV analysis\ldots}}

\begin{abstract}

The equivalence between the covariant and the non-covariant version of a
constrained system is shown to hold after quantization in the framework
of the field-antifield formalism.
Our study covers the
cases of Electromagnetism and Yang-Mills fields and sheds light on some
aspects
of the Faddeev-Popov method, for both the covariant and non-covariant
approaches, which have  not been fully clarified in the literature.

\end{abstract}


\clearpage

\section{Introduction}

\hspace{\parindent}%
In a recent paper \cite{OPT92} the BRST quantization of a class of
constrained dynamical systems was performed in the framework of the
Batalin-Vilkovisky (BV) formalism \cite{BV}.
These systems were specified by a Lagrangian which is quadratic in the
velocities and such that only primary first class constraints, linear in
the momenta, appear in its Hamiltonian analysis.
 After solving the Classical Master Equation --which is straightforward
due to
 its closed algebra structure-- the problem of the
ambiguity inherent to the resolution of the full Quantum Master Equation
was addressed. It is well known \cite{BV} that this ambiguity, which can
be drawn to the problem of defining the measure for the path integral, has
no solution within the BV formalism by itself and one has to rely on
other formulations --operator formalism, for instance-- to get the
correct answer. In this sense our result was promising: the physical
requirement
of making contact with the reduced path integral quantization procedure
--which is very close to ensuring unitarity-- is equivalent to the
internal
requirement (i.e. without departing from the BV formalism) of
choosing the solution of the Quantum Master Equation that makes the path
integral reparametrization invariant.

But a wide class of constrained systems do not fit within the theories
just considered. Many relevant physical examples, like
Electromagnetism (EM) and
Yang-Mills fields (YM), exhibit secondary as well as primary first-class
 Hamiltonian constraints.
There is an important physical reason for the appearance of secondary
constraints in these theories, and it is related to the way the
Hamiltonian formalism, which is manifestly non-covariant, is able to
provide us with gauge transformations which are Lorentz covariant.
Consider, for instance, the case of EM. The infinitesimal gauge
transformation
$\delta A_{\mu}=\derpar{\Lambda}{x^{\mu}}$ shows how to get a vector,
$\derpar{\Lambda}{x^{\mu}}$, from a scalar, $\Lambda$: just by taking
the gradient. The appearance of a time derivative of the arbitrary
function $\Lambda$ forces the Hamiltonian generator of the gauge
transformation to have two pieces, one with the first time derivative of
$\Lambda$ and the other one without a time derivative. Let us be
more specific;
we know on theoretical grounds \cite{Cast}, that a generator of gauge
transformations --acting through Poisson brackets-- must have the
following form
\be
G=\sum_{k = 0}^{N} G_{N-k} \Lambda^{k)},
\label{gautr}
\ee
$\Lambda^{k)}$ being the k-th time derivative of $\Lambda$ and $G_s$ an
s-generation first-class constraint.
In the case of EM, $G$ is the sum of two pieces, coming from one primary
and one secondary constraints (two generations). In fact, from the \break
Lagrangian ${\cal L} = -\frac{1}{4} F_{\mu \nu}F^{\mu \nu}$ we get a
canonical Hamiltonian \break
$H_c = \int d{\bf x} \left[\frac{1}{2}({\bf \pi}^2
+ {\bf B}^2) + {\bf \pi} \cdot \nabla A_0\right]$
and a primary constraint
(coming just from the definition of the momenta) $\pi_0 \simeq 0$.
Stability of this constraint under the Hamiltonian dynamics leads to
the secondary constraint $\dot\pi_0 = \{\pi_0, H_c\} = \nabla \cdot
{\bf \pi} \simeq 0$. No more constraints arise. Both the primary and the
secondary constraints are first-class and allow to write the gauge
generator (\ref{gautr}) in this case as
$$
G = \int d{\bf x} \left[\dot{\Lambda}({\bf x}, t) \pi^0({\bf x}, t)
+ \Lambda({\bf x},t) \nabla \cdot {\bf \pi}({\bf x}, t)\right] =
\int d{\bf x} \left(\derpar{\Lambda}{x^{\mu}} \pi^{\mu}\right),
$$
where $\Lambda$ is an infinitesimal arbitrary function. The gauge
transformation of the gauge field is then $\delta A_{\mu} =
\{A_{\mu}, G\} = \derpar{\Lambda}{x^{\mu}}$. We see therefore that a
primary and a secondary constraints are necessary to get
the gauge field $A_{\mu}$ transformed covariantly.

So, in principle, we are faced with the problem that the class of
theories studied
in \cite{OPT92}, which exhibit only primary constraints, seems  to exclude the
important physical case of covariant theories. In fact this is not true,
as we will see that our primary constraint theories can be conveniently
covariantized simply by promoting the Lagrangian multipliers associated
with the primary constraints to the status of dynamical variables,
covering  this way the cases of EM and YM fields ~\cite{fad}.
With this covariant theory at hand we can proceed to study the
new features that arise in this case and that were absent in the class
of systems studied in \cite{OPT92}, for instance the possibility of
having an open gauge algebra. This process of covariantization and the
study of the Hamiltonian and Lagrangian gauge algebra will be the topics
of the next section. In section 3 we perform the BV quantization of
the covariant theory by solving explicitly the Quantum Master Equation
and, after that, implementing several gauge fixings. In this fashion we
show the
equivalence between this covariant quantization and the non-covariant
approach used in \cite{OPT92}. Section 4 is devoted to conclusions.
Finally, an Appendix is introduced to show
how the covariantization procedure described in section 2 works in the
YM case.

\section{Gauge algebra and covariantization of the action}

\subsection{General setting}

\hspace{\parindent}%
The  non-covariant Lagrangians $L_{NC}(q,\dot{q})$%
\footnote{For the sake of simplicity we are going
to use throughout this paper the language of discrete degrees of
freedom. The switch to Field Theory language
can be done, at least formally, by using DeWitt's condensed notation.
Also for the same reason of simplicity we will restrict ourselves to the
case of classical bosonic variables, $\epsilon(q)=0$.}
of  interest to us are those for which
the tangent space is free from any constraints, yet $L_{NC}$
is still a singular Lagrangian. As  is proven in \cite{pap1}
this is equivalent to having in phase space only primary first-class
constraints. In this case, the canonical Hamiltonian $H_0(q,p)$ and
the constraints $T_{\alpha}(q,p)$, $\alpha = 1,...,r$, satisfy
\be
\{T_{\alpha},T_{\beta}\} = -C_{\alpha \beta}^{\gamma}(q,p)\,T_{\gamma},
\qquad
\{T_{\alpha},H_0\} = -V_{\alpha}^{\beta}(q,p)\,T_{\beta}.
\label{th}
\ee

Let us now have a quick look at the issue of ``covariantization''.
There is a simple way to get, from a canonical theory with only
primary first-class constraints, a classically
equivalent theory with primary
and secondary first-class constraints. This can be done by promoting
the Lagrangian multipliers $\lambda^{\alpha}$,
associated with the original constraints, to the
status of dynamical variables and to assume as a the new primary
constraints its canonical conjugate momenta $\pi_{\alpha}$.
Under such conditions the extended Hamiltonian will read
$$
   H(q,p;\lambda)=H_0(q,p)+\lambda^\alpha T_\alpha(q,p),
$$
and the stability of the new primary constraints
$\pi_{\alpha} \simeq 0$ will lead to the --now-- secondary constraints
$$
  \dot\pi_\alpha=\{\pi_\alpha, H\}=-T_\alpha(q,p)\simeq0,
$$
whose stability gives no new information
$$
  \dot T_\alpha=\{T_\alpha, H\}=-V^\beta_\alpha T_\beta
  -C^\gamma_{\alpha\beta}\lambda^\beta T_\gamma\simeq0.
$$

Using the well known algorithms briefly described in the
introduction \cite{Cast}
it is straightforward to construct the gauge generator for this case
$$
  G=\dot\veps^\alpha\pi_\alpha+\veps^\alpha
  \left[T_\alpha-(V^\beta_\alpha+C^\beta_{\alpha\gamma}\lambda^\gamma)
  \pi_\beta\right],
$$
and, consequently, the Hamiltonian gauge transformations for the
coordinates $q^A$, $\lambda^\alpha$
\bea
    \delta_H q^A&=&\{q^A,G\}=\{q^A,T_\alpha\}\veps^\alpha
    -\{q^A, (V^\beta_\alpha+C^\beta_{\alpha\gamma}\lambda^\gamma)\}
    \veps^\alpha\pi_\beta,
\nonumber\\
    \delta_H\lambda^\alpha&=& \{\lambda^\alpha, G\}=
    \dot\veps^\alpha-
    V^\alpha_\beta\veps^\beta-
    C^\alpha_{\beta\gamma}\lambda^\gamma\veps^\beta,
\label{gaugtr}
\eea
where $\veps^\alpha$ is an infinitesimal arbitrary function of time
(or space-time, in the case of field theory).

As is well known, to perform the covariant quantization of a gauge
theory within the framework of the field-antifield formalism, knowledge
of the gauge structure of the
classical Lagrangian theory is of fundamental importance. Since for the
systems under
consideration much of this information is
already contained in the Hamiltonian gauge structure, in what follows we
briefly describe the derivation of quantities and relations defining this
structure at the Hamiltonian level.

To begin with, let us derive
some relations involving the quantities $V^\alpha_\beta$,
$C^\alpha_{\beta\gamma}$, the constraints $T_\alpha$ and the
canonical hamiltonian $H_0$, which
appear as consequences of some Jacobi identities.
Consider, for instance, the following Jacobi identity
involving the constraints $T_\alpha$
$$
  \{T_\alpha,\{T_\beta,T_\gamma\}\}+
  \mbox{cyclic perm. of ($\alpha$, $\beta$, $\gamma$)}=0.
$$
Using \bref{th} we get
$$
  [C^\mu_{\sigma\gamma}C^\sigma_{\alpha\beta}-
  \{C^\mu_{\alpha\beta}, T_\gamma\}+
  \mbox{cyclic perm. of ($\alpha$, $\beta$, $\gamma$)}]\, T_\mu=0.
$$
The general solution of this equation
\be
  [C^\mu_{\sigma\gamma}C^\sigma_{\alpha\beta}-
  \{C^\mu_{\alpha\beta}, T_\gamma\}+
  \mbox{cyclic perm. of ($\alpha$, $\beta$, $\gamma$)}]=
  B^{\mu\rho}_{\alpha\beta\gamma}(q,p) \,T_\rho,
\label{a}
\ee
leads to the existence of a new function
$B^{\mu\rho}_{\alpha\beta\gamma}$, antisymmetric in its upper indices.

In much the same way, from the Jacobi identity among the constraints and
the canonical hamiltonian
$$
  \{T_\alpha,\{T_\beta,H_0\}\}+
  \{H_0,\{T_\alpha,T_\beta\}\}+
  \{T_\beta,\{H_0,T_\alpha\}\}=0,
$$
a new relation is obtained
\bea
  &&\left[-C^\sigma_{\gamma\beta} V^\gamma_\alpha+
  C^\sigma_{\gamma\alpha} V^\gamma_\beta+
  C^\gamma_{\alpha\beta} V^\sigma_\gamma\right.
\nonumber\\
  &&\left.+\{V^\sigma_\alpha, T_\beta\}-\{V^\sigma_\beta,
   T_\alpha\}-\{C^\sigma_{\alpha\beta}, H_0\}\right]=
   D^{\sigma\mu}_{\alpha\beta}(q,p) T_\mu,
\label{b}
\eea
yielding the appearence of
a new structure function, $D^{\sigma\mu}_{\alpha\beta}$,
antisymmetric in its upper and lower indices.

Continuing this procedure, that is, taking an increasing number of
Poisson brackets among the constraints and the canonical Hamiltonian and
antisymmetrizing them in a convenient fashion,
new quantities and new relations among the functions previously
obtained are found. All these objects are the so-called
structure functions and the relations among them determine the structure
of the Hamiltonian gauge algebra. For a more exhaustive study of this
Hamiltonian gauge structure we refer the reader to ref.\cite{Hen}.

\subsection{The model: Quadratic Lagrangians}

\hspace{\parindent}%
So far we have established the most general setting for theories we
are interested in. Now we will apply this framework
to the case of non-covariant quadratic Lagrangians of the type
\cite{OPT92}
\be
    L_{NC}(q,\dot q)=\frac{1}{2}\,\dot q^A G_{AB}(q) \dot q^B-
    V(q),\quad\quad
    A,B=1,\ldots,N,
\label{lagrang}
\ee
where $G_{AB}(q)$ is a singular metric such that its null vectors
$U^A_{\alpha}(q)$, $G_{AB}U^B_{\alpha} = 0$, $\alpha = 1,\ldots,k,$ are
Killing vectors for it, i.e.
\be
   ({\cal L}_\alpha G)_{AB}= G_{AB,C} U^C_\alpha+
    G_{AC} U^C_{\alpha,B}+G_{BC} U^C_{\alpha,A}=0,
\label{kill}
\ee
and keep the potential $V$ unchanged
$$
U_{\alpha}^A \derpar{V}{q^A} = 0.
$$
In \bref{kill}, ${\cal L}_{\alpha}$ stands for
the Lie derivative in the $\hat{U}_{\alpha}$ direction.
These last two conditions enforce the non-existence of Lagrangian
constraints \cite{OPT92}.

The primary Hamiltonian constraints for this system are easily found
\be
   T_{\alpha} = U_{\alpha}^A p_A,
\label{primary}
\ee
and its first-class character, which is guaranteed by requirement
\bref{kill} yields the commutation relations defining the
structure functions $C^\gamma_{\alpha\beta}$
\be
  U^A_{\alpha,B}U^B_\beta- U^A_{\beta,B}U^B_\alpha=
  -U^A_\gamma C^\gamma_{\alpha\beta},
\label{struc func}
\ee
which in the present case depend only on the coordinates $q^A$.

On the other hand, the canonical Hamiltonian $H_0$ associated to
(\ref{lagrang}) is
\be
  H_0(q,p)= \frac{1}{2} p_A M^{AB}(q) p_B + V(q),
\label{ham}
\ee
where the metric
$M^{AB}$ is a symmetric non-singular matrix satisfying
\be
M^{AB}G_{AC}G_{BD} = G_{CD}.
\label{mgg}
\ee

Let us notice that the metric
$M^{AB}$ defined through (\ref{mgg}) displays a certain degree of
arbitrariness. This corresponds to the fact that the canonical
Hamiltonian is only unambiguously defined on the primary constraint
surface%
\footnote{As is proven in \cite{pap1} and \cite{pap2}, this
arbitrariness has no effect when a reduced (classical elimination of the
gauge degrees of freedom) quantization is performed.}.

Taking into account (\ref{mgg}) and the fact that the vector fields
$\hat{U}_{\alpha} = U_{\alpha}^A \derpar{}{q^A}$
are Killing vectors for the metric $G$, \bref{kill}, we immediately
obtain
$$
({\cal L}_{\alpha} M)^{AB}G_{AC}G_{BD} = 0.
$$
This result leads to the following form for $({\cal L}_{\alpha} M)$
\be
({\cal L}_{\alpha} M)^{AB}= U_{\beta}^A M_{\alpha}^{\beta B}(q) +
 U_{\beta}^B M_{\alpha}^{\beta A}(q).
\label{lieder}
\ee
Notice that the choice of $M_{\alpha}^{\beta A}(q)$ is again ambiguous.
In fact there is a family of such  possible
objects, related to each other by
\be
M_{\alpha}^{\beta A} \rightarrow M_{\alpha}^{'\beta A} =
M_{\alpha}^{\beta A} + G_{\alpha}^{\beta \gamma} U_{\gamma}^A,
\label{closing}
\ee
$G_{\alpha}^{\beta \gamma}$ being
an arbitrary array of coefficients antisymmetric in its upper indices. In
the next subsection we will take advantage of this arbitrariness.

{}From the above results the form of the structure
functions $V^\alpha_\beta$ is easily worked out. Indeed, taking into account
its definition, the form of the constraints $T_\alpha$ \bref{primary} and the
Hamiltonian $H_0$ \bref{ham} we have
$$
\{T_{\alpha},H_0\} = -\frac{1}{2} p_A ({\cal L}_{\alpha} M)^{AB} p_B =
- V_{\alpha}^{\beta} T_{\beta},
$$
where use of the relation \bref{lieder} allows to factorize the
constraints and write the structure functions $V_{\alpha}^{\beta}$ as
\be
  V_{\alpha}^{\beta}(q,p) = M_{\alpha}^{\beta A}(q) p_A.
\label{ve}
\ee

Finally, let us write down the consequences of the Jacobi identities for
the constraints and the canonical Hamiltonian in our particular model.
{}From relation \bref{a} we obtain
$$
C_{\sigma \gamma}^{\mu} C_{\alpha \beta}^{\sigma} -
C_{\alpha \beta}^{\mu},_A U^A_{\gamma} +
\mbox{cyclic perm. of}\,(\alpha,\beta,\gamma) = 0,
$$
the structure functions $B^{\mu\rho}_{\alpha\beta\gamma}$ vanishing in
this case, due to the fact that $C^\gamma_{\alpha\beta}$ and
$U^A_\alpha$ depend only on $q^A$. On the other hand
\bref{b} turns out to be
\bea
  &&\left[-C^\sigma_{\gamma\beta} V^\gamma_\alpha+
  C^\sigma_{\gamma\alpha} V^\gamma_\beta+
  C^\gamma_{\alpha\beta} V^\sigma_\gamma\right.
\nonumber\\
  &&\left.+\{V^\sigma_\alpha, T_\beta\}-\{V^\sigma_\beta, T_\alpha\}
  -C^\sigma_{\alpha\beta,A}M^{AB}p_B\right]=
   D^{\sigma\mu}_{\alpha\beta} T_\mu,
\label{jacobi 4}
\eea
where now the structure functions $D^{\sigma\mu}_{\alpha\beta}$
can be chosen to depend only on $q^A$, as it is seen if the linear
dependence of the
constraints $T_\alpha$ and the structure functions $V^\alpha_\beta$ on
$p_A$ is taken into account.

This analysis could be carried on to determine the higher order
structure functions. Nevertheless, since these quantities will not
appear in the situation we will consider, we do not  pursue
this direction any further. Rather, in what follows, we are going to
undertake the study of the Lagrangian gauge structure using as
background the above results.

\subsection{Covariantization and Lagrangian gauge structure}

\hspace{\parindent}%
Having studied the Hamiltonian gauge algebra
we are ready for ``covariantization''.
Using the Lagrangian multipliers as new variables,
the extended Hamiltonian reads in our particular case
$$
   H(q,p;\lambda)=H_0(q,p)+\lambda^\alpha T_\alpha(q,p)=
   \frac{1}{2} p_A M^{AB}(q) p_B + V(q)+\lambda^\alpha U^A_\alpha p_A.
$$
To obtain the associated Lagrangian we should eliminate the momenta
$p_A$ in terms of the velocities $\dot q^A$. Use of the equations
of motion yields
$$
  \dot q^A=\der{H}{p_A}= M^{AB} p_B + U^A_\alpha\lambda^\alpha,
$$
and since $M^{AB}$ is invertible, we have
$$
   p_A(q,\dot q;\lambda)=M_{AB}(\dot q^B- U^B_\beta\lambda^\beta),
$$
with $M_{AB}M^{BC} = \delta^C_A$.

The corresponding ``covariant'' Lagrangian $L_C$ is
\be
  L_{C}(q,\dot q;\lambda)= \frac12 (\dot q^A- U^A_\alpha\lambda^\alpha)
  M_{AB}(q) (\dot q^B- U^B_\beta\lambda^\beta)-V(q).
\label{nou lag}
\ee
In the Appendix we show that in the case of Yang-Mills theories
(\ref{nou lag}) is the standard covariant Lagrangian for these
systems.

This Lagrangian $L_C(q,\dot{q};\lambda)$
will be the starting point
of our analysis. First we can check that the pull-back of the
transformations (\ref{gaugtr}), given by
\bea
    \delta q^A&=&U^A_\alpha(q)\veps^\alpha,
\nonumber\\
    \delta\lambda^\alpha&=&\dot\veps^\alpha-
    M^{\alpha A}_\beta p_A(q,\dot q;\lambda) \veps^\beta-
    C^\alpha_{\beta\gamma}(q)\lambda^\gamma\veps^\beta,
\label{gaugtr2}
\eea
are indeed, as was expected, gauge transformations for $L_C$
$$
\delta L_C = 0.
$$

As we have said, the structure of the
algebra of the Lagrangian gauge transformations plays a crucial role in
solving the Master Equation in the field-antifield approach --which
is the subject of the next section. In our case, for
$\delta_1 := \delta[\veps_1]$, $\delta_2 := \delta[\veps_2]$,
we obtain
$$
[\delta_1, \delta_2]q^A =
\delta[C_{\alpha \beta}^{\gamma} \veps_2^{\beta}\veps_1^{\alpha}]q^A,
$$
and, after a lengthy calculation
\bea
[\delta_1, \delta_2]\lambda^\alpha&=&
\delta[C_{\sigma\beta}^{\gamma} \veps_2^{\beta} \veps_1^{\sigma}]
\lambda^\alpha
\nonumber\\
   &+&\left\{[M^{\alpha A}_\mu M_{AB} M^{\sigma B}_\beta-
     (\mu\leftrightarrow\beta)]+D^{\alpha\sigma}_{\beta\mu}\right\}
    \der{L_C}{\lambda^\sigma}\veps^\mu_1\veps^\beta_2,
\label{open}
\eea
where $D^{\alpha\sigma}_{\beta\mu}$ are the (pull-back of the)
Hamiltonian
structure functions defined through relation \bref{jacobi 4} and
$\derpar{L_C}{\lambda^\alpha}$ the equations of motion for the fields
$\lambda^\alpha$ derived from the covariant Lagrangian $L_C$
\bref{nou lag}, given by
\be
    \der{L_C}{\lambda^\alpha}=-U^A_\alpha M_{AB}(\dot q^B-U^B_\beta
    \lambda^\beta).
\label{lambda eq}
\ee

In the study of this gauge algebra we meet for the first time the new
features introduced in the theory by the process of covariantization.
Indeed,
observe that the structure of (\ref{open}) is, in general, that of
an {\it open algebra}. This fact makes the computation of the
proper solution of the Master Equation rather cumbersome and we will
try to circumvent this problem.
To this end we will use a result from ref.\cite{BV85}, to wit:
{\it any open algebra of gauge transformations may be closed by
the addition of
the appropiate antisymmetric combinations%
\footnote{It should be noted that in a general case with both bosons and
fermions, these combinations will have a graded antisymmetry.}
of the equations of motion}.
In our specific case, since the openness of the algebra only concerns
the $\lambda$ sector and, moreover, its open algebra part
-see (\ref{open})-
only exhibits the equations of motion for the $\lambda$'s,
$[L_C]_{\lambda^{\alpha}} = \derpar{L_C}{\lambda^{\alpha}}$,
\bref{lambda eq}, it seems
to be very {\it plausible} that we can get the closed algebra structure
just by leaving $\delta{q^A}$ unchanged and
modifying $\delta \lambda^{\alpha}$ (\ref{gaugtr2}) as
follows
$$
\delta \lambda^{\alpha} \rightarrow \delta' \lambda^{\alpha} =
\delta \lambda^{\alpha} +
F^{\alpha \beta}_{\gamma} \derpar{L_C}{\lambda^\beta}\veps^\gamma,
$$
with an appropriate $F^{\alpha\beta}_\gamma$ antisymmetric in its upper
indices.
Using (\ref{gaugtr2}) and the explicit expression for the equations of
motion of $\lambda$, \bref{lambda eq}, this can also be written as
$$
  \delta'\lambda^{\alpha}= \dot\veps^\alpha-
   M^{'\alpha A}_\beta p_A(q,\dot q;\lambda) \veps^\beta-
    C^\alpha_{\beta\gamma}(q)\lambda^\gamma\veps^\beta,
$$
with
$$
   M^{'\alpha A}_\beta=M^{\alpha A}_\beta +
   F^{\alpha\gamma}_\beta U^A_\gamma.
$$

This last expression simply displays the freedom in the choice
of $M_{\alpha}^{\beta A}$ we discovered in (\ref{closing}). We can
therefore conclude that it is plausible that
the freedom described by (\ref{closing})
allows for a choice of  $\delta' \lambda^{\alpha}$ which satisfies
the closed algebra structure. Strictly speaking we have not proven
this, although it is very plausible, as we have argued.

Actually we may have considered from the beginning a more restrictive
case: the assumption \cite{DW}, for instance, that the regular metric
tensor $M^{AB}$ be such that the action of the gauge group leads to
isometries, i.e.
$({\cal L}_{\alpha} M)^{AB} = 0, \alpha =1, \ldots ,k$. In fact, this
Killing condition implies that the vector fields $\hat{U}_{\alpha}$ form
a Lie algebra ($C_{\alpha \beta}^{\gamma}$ = constant). Indeed,
eqs.(\ref{struc func}) with $ C_{\alpha \beta}^{\gamma}$ = constant, are
the integrability conditions for
$({\cal L}_{\alpha} M)^{AB} = 0$, $\alpha =1,\ldots ,k$.
In that case, the treatment of the system greatly simplifies:
from (\ref{lieder}) we see that the quantities $M^{\beta B}_{\alpha}$
and, as a consequence, the structure functions $V^{\beta}_{\alpha}$ of
(\ref{ve}) can be chosen to be zero. Then
$D_{\alpha \beta}^{\sigma \mu}$ defined in (\ref{jacobi 4}) can be put
to zero
as well. All these results together lead to the closure of the gauge
algebra of (\ref{open}) in this particular case. Note that the important
cases of EM and YM, for which the structure functions
$C^\gamma_{\alpha\beta}$ are constants, fall into this last category and
have a closed gauge algebra.

{}From now on, whatever be the case we are dealing with, we will assume
that
we have met the conditions to get the gauge algebra in the closed form
$$
[\delta_1, \delta_2](q^A, \lambda^\alpha) =
\delta[C_{\alpha \beta}^{\gamma} \veps_2^{\beta}
\veps_1^{\alpha}](q^A,\lambda^\alpha).
$$
This assumption greatly simplifies the determination of the solutions
of the Classical and Quantum Master Equations, which we are going to
undertake in the next section.

\section{BV quantization of the covariant action}

\hspace{\parindent}%
In the case of an irreducible closed algebra of gauge transformations,
the field-antifield formalism starts by enlarging the original
configuration space with the introduction of a ghost $c^{\alpha}$
for each gauge parameter $\veps^\alpha$, of opposite parity. These
ghosts, together with the classical fields $\{\phi^a\}=\{q^A,
\lambda^{\alpha}\}$, form the minimal sector of fields $\{\phi^i\}$.
A new set of ``antifields'',
$\{ q^*_A, \lambda^*_{\alpha}, c^*_{\alpha}\}=\{\phi^*_i\}$,
with  parities opposite to those of its associated fields, is introduced as
well. Then, in the space of functionals of the fields and their antifields,
some new structures, the antibracket
$$
(A,B) =\rder{A}{\phi^i}\lder{B}{\phi^*_i}-
       \rder{A}{\phi^*_i}\lder{B}{\phi^i},
$$
and the BRST ``Laplacian''
$$
  \Delta=\rder{}{\phi^i}\lder{}{\phi^*_i},
$$
are defined (sum over continuous indices is understood in both
structures).
The Quantum Master Equation is then formulated as an equation
for a functional $W$ -the full quantum action-
\be
(W,W) - 2i \hbar \Delta W =0.
\label{qme}
\ee

The usual way to solve the above equation consists in expanding the
quantum action $W$ in powers of $\hbar$,
$$
W=S + \sum_{m=1}^{\infty} {\hbar}^m W_m,
$$
so that (\ref{qme}) splits into the Classical Master Equation
\be
(S,S) = 0,
\label{cme}
\ee
and the equations for the "quantum corrections"
\bea
(W_1,S)&=& i\Delta S,
\nonumber\\
(W_p,S)&=& i\Delta W_{p-1} - \frac{1}{2}
\sum_{q=1}^{p-1}(W_q,W_{p-q}), \quad  p \geq 2.
\label{q2}
\eea

For an irreducible, closed gauge algebra
the Classical Master Equation (\ref{cme})
has the well known (minimal) proper solution
$$
  S= S_0(\phi)+\phi^*_a R^a_\alpha c^\alpha
  +\frac12 c^*_\alpha T^\alpha_{\beta\gamma} c^\gamma c^\beta,
$$
where $S_0$, $R^a_\alpha$ and $T^\alpha_{\beta\gamma}$ are the classical
action, the generators of the gauge transformations and the structure
functions of the gauge algebra, respectively. In our case, taking into
account the results obtained in the preceding section and assuming to
have met the conditions to get the algebra in closed form, we obtain the
following expression for the proper solution
$$
  S= S_C(q,\lambda)+q^*_A U^A_\alpha c^\alpha
    +\lambda^*_\alpha(\dot c^\alpha-V^\alpha_\beta c^\beta-
     C^\alpha_{\beta\gamma} \lambda^\gamma c^\beta)
  -\frac12 c^*_\alpha C^\alpha_{\beta\gamma} c^\gamma c^\beta,
$$
where now $S_C$ is the classical
action associated with the covariant Lagrangian \bref{nou lag},
$U^A_\alpha(q)$ the gauge generators for the fields $q^A$ and
$V^\alpha_\beta$, $C^\alpha_{\beta\gamma}$ the pull back of the
corresponding Hamiltonian structure functions \bref{struc func} and
\bref{ve}.

Let us consider now the equations for the quantum corrections \bref{q2}.
Since for the type of theories we are analyzing the proper
solution is linear in the antifields, the quantity $\Delta S$ does not
depend on them. As a consequence, the term $W_1$ can be chosen to be a
function of the classical fields only and eqs.\bref{q2} are immediately
solved by taking $W_p=0$ for $p\geq2$. Therefore, no higher order terms
in $\hbar$ appear apart from the $W_1$ term.

To compute the first quantum correction $W_1$ we need
the explicit expression of $\Delta S$. In our case it is easily seen
that
$$
\Delta S = (U^A_{\alpha,A}
           +M^{\beta A}_\alpha M_{AB} U^B_\beta) c^\alpha.
$$
Use of the Lie derivative of $M^{AB}$ in the $\hat U_\alpha$
direction \bref{lieder} together with the symmetry property of
this metric allows to write the last term of the above expression as
\bea
  &&\left(M^{\beta A}_\alpha M_{AB} U^B_\beta\right) c^\alpha=
  \left[\frac12 M_{AB}(\gh L_\alpha M)^{AB}\right] c^\alpha=
\nonumber\\
  &&\left[-\frac12 M^{AB}(\gh L_\alpha M)_{AB}\right] c^\alpha=
  -\left(\frac12 M^{AB} M_{AB,C} U^C_\alpha+U^A_{\alpha,A}\right)
  c^\alpha,
\nonumber
\eea
where in the last equation use has been made of the definition of
$(\gh L_\alpha M)_{AB}$. Therefore, after these manipulations, $\Delta
S$ can be written in the more useful form
\bea
\Delta S&=&  \left[ -\frac{1}{2} \hat{U}_{\alpha}
(\tr\ln M_{AB}) c^{\alpha} \right]
\nonumber\\
  &=&\left[ -\frac{1}{2} \hat{U}_{\alpha}
(\ln\det M_{AB}) c^{\alpha} \right]
= -\left(\ln(\det M_{AB})^{1/2}, S\right).
\label{deltas}
\eea
Notice that the trace over continuous indices imply, in our case
of a local gauge theory, that $\Delta S$ is proportional to $\delta(0)$.
Therefore, in order to make sense out of this construction, some scheme
to regularize the above expression must be considered.

Expression \bref{deltas} is already spelling out the formal solution for
$W_1$. Indeed, we can simply take
\be
W_1  = -i \ln(\det M_{AB})^{1/2} +\mbox{BRST-invariant terms},
\label{w1}
\ee
where by ``BRST-invariant terms'' we mean terms with vanishing
antibracket with $S$. As we have said,
the above ambiguity in $W_1$ is inherent to the field-antifield
formalism. In the present case
we solve this ambiguity just by dropping the second term in the lhs of
\bref{w1}. As  will be shown below, this is the correct choice
that makes contact with the reduced path integral formalism.

Now, to proceed to fix the gauge within the field-antifield
approach, some auxiliary fields,
$\left\{\bar{c}^{\alpha}, B^{\alpha}\right\}$, and its corresponding
antifields, $\left\{\bar{c}^*_{\alpha}, B^*_{\alpha}\right\}$,
are introduced. After that, the minimal proper solution $S$
is modified by the addition of an extra term in these new fields as
$$
  S_{\rm n.m.}=S+\bar{c}^*_\alpha B^\alpha.
$$
Then, if we call $\Phi^i$
the whole set of fields, a "gauge fermion" $\Psi$ will eliminate the
antifields through the requirement
$$
\Phi^*_i = \derpar{\Psi}{\Phi^i}.
$$

The Batalin-Vilkovisky path integral is then defined as
\be
Z_{\Psi} = \int [Dq][D \lambda][D \bar{c}][Dc][DB]
\exp\left\{\frac{i}{\hbar} W_{\Sigma}\right\},
\label{bpi}
\ee
where $W_{\Sigma}$ stands for $W(\Phi, \Phi^* = \derpar{\Psi}{\Phi})$.
It should be noted that, in our case, as $W_1$ does not have any
dependence on the antifields $\Phi^*$, its expression will not depend on
the choice of the gauge fixing. Therefore, we will have
$$
  W_{\Sigma}=S(\Phi, \Phi^* = \derpar{\Psi}{\Phi})+\hbar W_1(\Phi).
$$

Now, a customary lattice regularization for
$\delta(0)$, $\delta(0) \rightarrow \frac{1}{\veps},
\veps \rightarrow 0 $, allows us to write part of the exponential
in \bref{bpi} as
$$
  \exp\left\{i W_1\right\} =  \prod_t (\det M_{AB})^{1/2},
$$
so that $Z_{\Psi}$ becomes
\be
Z_{\Psi} = \int [Dq][D \lambda][D \bar{c}][Dc][DB]
(\det M_{AB})^{1/2}
\exp\left\{\frac{i}{\hbar} S_{\Sigma}\right\}.
\label{zeta or}
\ee
Therefore, from the above expression for $Z_\Psi$ it is evident that
while the gauge fixed proper solution of the classical master equation
\bref{cme} represents the classical effective action of the theory, the
$W_1$ term stands for quantum corrections to the naive measure. In this
way, the determinant $(\det M_{AB})^{1/2}$ modifies the naive measure
$[Dq]\ldots[DB]$ yielding a BRST invariant measure.

At this point, different choices of the gauge fixing fermion
are possible. The physical equivalence of the different gauges, i.e.
the invariance of the path integral \bref{bpi} under deformations
of the gauge fixing fermion,
is a well known result in the context of the field-antifield
formalism, and has been proven by Batalin and
Vilkovisky in \cite{BV}. In fact, what they do in this reference is to
prove the theorem for gauges that differ infinitesimally. More suited
to our purposes, a direct proof of this invariance under arbitrary
deformations of $\Psi$ for the case of theories with closed, irreducible
gauge algebras, can be found in ref. \cite{B81}
and will not be repeated here.
{}From now on, we will take for granted this invariance of the
path integral \bref{bpi} under changes of the gauge fixing fermion%
\footnote{This will be true as long as the theory is free from gauge
anomalies. We assume that this is the case in this paper. See however
ref.\cite{VanN}.}.

One of the main purposes of this paper is
to make contact with the non-covariant path integral quantization
presented in \cite{OPT92}. To this end, we will use a gauge fixing
fermion implementing unitary or, more generally,
non-covariant gauge fixing conditions. However,
other types of gauge fixing conditions,  for instance ``covariant''
gauges, can be chosen as well. In what follows, we
are going to work out
the form of the Batalin-Vilkovisky path integral in both classes of
gauges.

Unitary or non-covariant gauge conditions are necessary
in order to make contact with the reduced path integral
quantization. In this formulation, unitarity is obvious once
the usual assumptions about the positivity of the spectrum of the
reduced theory are made. Gauge fixing fermions which implement these
gauges are taken to be of the form
$\Psi_1 = \bar{c}_{\alpha} F^{\alpha}(q)$, where the gauge fixing
conditions $F^\alpha(q)$ do not involve the Lagrange multipliers
$\lambda^\alpha$. For such gauge fixing fermions $S_{\Sigma_1}$ becomes
$$
 S_{\Sigma_1} = S_C + \bar{c}_{\alpha}\der{F^{\alpha}}{\veps^{\beta}}
c^{\beta} + B_{\alpha}F^{\alpha},
$$
where
$\derpar{F^{\alpha}}{\veps^{\beta}} =\hat{U}_{\beta}(F^{\alpha})$
measures the rate of change of $F^{\alpha}$ under the action of the gauge
generators $\hat{U}_ {\beta}$. After this, straightforward integration
of $c, \bar{c}$ and $B$ in \bref{zeta or} yields
\be
  Z_{\Psi_1} = \int [Dq][D \lambda]
  (\det M_{AB})^{1/2}\,\det\left(\der{F^\alpha}{\eps^\beta}\right)\,
  \delta(F^\alpha)\,\exp\left\{\frac{i}{\hbar} S_C\right\}.
\label{fc}
\ee

The non-covariant formulation is recovered by integrating out
the Lagrange multipliers $\lambda$, which appear quadratically in $S_C$.
Once this is done we get
\be
  Z_{\Psi_1} = \int [Dq]
  \frac{(\det M_{AB})^{1/2}}{(\det \theta_{\alpha\beta})^{1/2}}
\,\det\left(\der{F^\alpha}{\eps^\beta}\right)\,
  \delta(F^\alpha)\,\exp\left\{\frac{i}{\hbar} S'_0\right\},
\label{znc}
\ee
where $\theta_{\alpha\beta}$ is defined by
$$
   \theta_{\alpha\beta}=U^A_\alpha M_{AB} U^B_\beta,
$$
and the new classical action $S'_0$ is
$$
  S'_0=\int\dif t\,\left\{\frac12 \dot q^A M_{AB} \dot q^B-\frac12
  (\dot q^A M_{AC} U^C_\alpha)(\theta^{-1})^{\alpha\beta}
  (\dot q^B M_{BD} U^D_\beta)+ V(q)\right\}.
$$
It can be shown that the metric defining the kinetic term of $S'_0$ is
nothing but the one which appears in the original quadratic
non-covariant Lagrangian \bref{lagrang}, i.e.
$$
   G_{AB}=M_{AB}- M_{AC}U^C_\alpha(\theta^{-1})^{\alpha\beta}
                  M_{BD} U^D_\beta.
$$
Therefore, we conclude that $S'_0$ is equal to $S_{NC}$ --the action for
the non-covariant Lagrangian \bref{lagrang}-- and that expression
\bref{znc} for $Z_{\Psi_1}$ in
non-covariant gauges is in complete agreement with the one obtained in
\cite{OPT92}.
This result proves that the equivalence between the non-covariant
(with variables $q$ only) and the covariant (with variables $q$ and
$\lambda$)
formulations, which is easily seen at the classical level, still holds
after quantization of the theory within the framework of the
field-antifield formalism when a non-covariant gauge fixing is imposed.

At this point it is worth comparing the two versions, \bref{fc} and
\bref{znc}, of the path integral $Z_{\Psi_1}$. In fact, they are two
different, although equivalent, Faddeev-Popov (FP) formulas. On the one
hand, expression \bref{fc} for
the covariant theory corresponds to the standard FP formula as used in
the literature (with two extensions: the presence of a non-trivial
determinant in the measure and also the fact that we are dealing with
the so-called quasigroup structure \cite{B81} rather than a Lie group).
On the other hand, the
equivalent expression \bref{znc} uses a non-covariant action and
corresponds to the correct FP formula for systems with first-class primary
constraints only (strictly speaking, systems with quadratic kinetic term
and constraints linear in the momenta), as it was proven in \cite{pap2}.
In this second case, it should be noted the presence of a new
determinant, $(\det\theta_{\alpha\beta})^{1/2}$, which makes the path
integral invariant under rescaling of the constraints. In summary, the
above discussion
points out that {\it the structure of the constraints of the theory
(primary
constraints in \bref{znc}; primary and secondary in \bref{fc}) makes a
difference with regard to the final form of the FP formula.}

In connection with the measure of our path integral \bref{zeta or}
another comment is in order. As we have said in the preceding section,
there is a certain amount of arbitrariness in the selection of the
metric $M_{AB}$ fulfilling \bref{mgg}. One may then wonder how this
arbitrariness affects the path integral \bref{zeta or}. In fact,
using expression \bref{znc} of $Z_\Psi$ in a non-covariant gauge, one
can see that it does not affect it at all. This expression was obtained
in \cite{pap2} starting from the reduced path integral quantization,
in which this kind of ambiguity was not present%
\footnote{See footnote 2 in relation with the effects of this
arbitrariness at the quantum level.}.
Therefore, in spite
of this apparent dependence of \bref{zeta or} on the particular choice
of $M_{AB}$, the measure and the action depend on it in such a way  that
\bref{zeta or}, in the end, does not suffer from this arbitrariness.
In the measure, this feature is neatly displayed as a cancellation of
the dependence on the gauge part of $(\det M_{AB})^{1/2}$,  namely,
$(\det M_{\alpha\beta})^{1/2}$, and the similar
dependence in $(\det \theta_{\alpha\beta})^{1/2}$ ( in \cite{pap2} it was
shown that $(\det M_{AB})^{1/2}$  factorizes into
a physical piece - dependent only on gauge invariant degrees of freedom -
times $(\det M_{\alpha\beta})^{1/2}$ . This was achieved in the so-called
adapted coordinates; the
gauge dependent piece of $(\det \theta_{\alpha\beta})^{1/2}$
is also explicitely displayed this way).

To conclude, for the sake of completeness, let us study the form of
$Z_\Psi$ \bref{zeta or} in covariant gauges. As is well known,
covariant gauge fixings are more convenient in
obtaining Feynman rules which describe the perturbative sector of the
quantum theory. This class of gauges is constructed so that all the
fields become propagating. Gauge fixing fermions enforcing covariant
gauges are usually of the form
$$
  \Psi_2= \bar c^\alpha\left[\dot\lambda^\alpha+F(q)+\frac12
  \omega^{\alpha\beta} B_\beta\right],
$$
where the maximum rank metric $\omega^{\alpha\beta}$ is usually taken to
be independent of the fields. The gauge fixed action $S_{\Sigma_2}$
reads in this case
$$
 S_{\Sigma_2} = S_C + \bar{c}_{\alpha}\der{F^{\alpha}}{\veps^{\beta}}
c^{\beta} -\dot{\bar c}_\alpha\left(\dot c^\alpha-V_\beta^\alpha c^\beta
  -C^\alpha_{\beta\gamma}\lambda^\gamma c^\beta\right)
   +B_\alpha  \left(\dot\lambda^\alpha+F(q)+\frac12
  \omega^{\alpha\beta} B_\beta\right).
$$
The auxiliary fields $B_\alpha$ can be integrated out of the
path integral or, equivalently, eliminated algebraically in terms of
their equations of motion
$$
   \der{L_{\Sigma_2}}{B_\alpha}=    \dot\lambda^\alpha+F(q)+
  \omega^{\alpha\beta} B_\beta=0\Rightarrow
  B_\alpha=-\omega_{\alpha\beta}(\dot\lambda^\beta+F^\beta(q)),
$$
where $\omega_{\alpha\beta}$ is the inverse of the metric
$\omega^{\alpha\beta}$, yielding in this way a gauge fixed action of the
form
\bea
 S_{\Sigma_2} &=& S_C + \bar{c}_{\alpha}\der{F^{\alpha}}{\veps^{\beta}}
c^{\beta} -\dot{\bar c}_\alpha\left(\dot c^\alpha-V_\beta^\alpha c^\beta
  -C^\alpha_{\beta\gamma}\lambda^\gamma c^\beta\right)
\nonumber\\
   &&+\frac12(\dot\lambda^\alpha+F^\alpha(q))  \omega_{\alpha\beta}
   (\dot\lambda^\beta+F^\beta(q)),
\nonumber
\eea
in which the kinetic terms of all the fields
are invertible, so that they become propagating. This is an important
feature which distinguish the covariant formulation (i.e. with variables
$\lambda$) from the non-covariant one.

Finally, the partition function in covariant gauges
is written as
$$
Z_{\Psi_2} = \int [Dq][D \lambda][D \bar{c}][Dc]
(\det M_{AB})^{1/2}
\exp\left\{\frac{i}{\hbar} S_{\Sigma_2}\right\},
$$
this expression being the starting point in the construction of
covariant Feynman rules.

\section{Conclusions}

\hspace{\parindent}%
In this paper we have extended to Yang-Mills type systems some previous
work on the quantization of constrained systems which exhibited, in the
canonical formalism, only primary first-class constraints linear in the
momenta. This
extension can be understood as the covariantization of the original
system by introducing new degrees of freedom to it.
At this point it is worth noticing that the
marriage of covariance (for a constrained system like YM, for instance)
with the Hamiltonian formalism immediately implies the appearance of
secondary constraints. Due to this fact, there are some differences
in the application of the Batalin-Vilkovisky formalism to both the
non-covariant
and the covariant case which deserve some specific comments.
The main difference is perhaps that the algebra of gauge transformations
will be generally open in the covariant case, even though it was closed
in the
non-covariant one. In this paper we have dealt with this eventuality
by
arguing that it is possible to set up the covariant formalism in such a way
that the algebra is still closed, and this is in fact the only case we
have studied and where the equivalence with the non-covariant
formulation has been shown.

Another difference, which can be traced to the different structure of
the constraints in both cases, is the following:
 in the non-covariant case (which in our
terminology corresponds to a system without Lagrangian constraints or,
in other words \cite{pap1}, with only primary first-class
Hamiltonian constraints), there appears \cite{pap2}
in the Faddeev-Popov formula a new determinant, unrelated to the gauge
fixing procedure, that keeps the path integral invariant under
rescaling of
the constraints (which we emphasize are linear in the
momenta). Instead, in the covariant case (which is achieved by promoting
the old Lagrangian multipliers to the status of dynamical variables,
thus creating two generations, primary and secondary, of constraints),
the new determinant is absorbed in the definition of the covariant
Lagrangian, and the usual Faddeev-Popov formula is obtained. Our result,
however,
is an extension of the Faddeev-Popov formula because now
the generators of the gauge group do not span a Lie algebra. The
structure defined by these generators --whose commutation relations give
rise to structure functions, unlike the structure constants that appear
in a Lie algebra-- has been called a quasigroup \cite{B81}.

In conclusion, our results establish the equivalence, at the quantum
level, of the
non-covariant and the covariant version of a constrained dynamical system
of Yang-Mills type. This equivalence is a fundamental issue because, in
terms of path integrals, unitarity is best checked in the
reduced quantization (classical elimination of the gauge degrees of
freedom). This reduced quantization corresponds, as it is proven
in \cite{pap2}, to the quantization of the non-covariant version of the
system.

\medskip

During the preparation stages of this manuscript we received a preprint
by Epp et al. \cite{Kun}, which deals with some of the  topics
raised here,
as well as with other work by some of us. We completely agree with
their results. The crucial point first raised in \cite{pap2}, and clarified
to some extent in \cite{Kun} using scalar QED as an example, is the need
to distinguish between different forms of the Faddeev-Popov ansatz when both
primary and secondary constraints are present classically (i.e. before the
Lagrange multipliers are integrated out) and when only primary constraints
are present (i.e. after the Lagrange multipliers are integrated out). Thus
for example the usual, covariant form of the Faddeev-Popov ansatz, ( and
consequently formula (3.31) of \cite{Ep} is correct only when secondary
constraints are present. Ref. \cite{pap2} dealt specifically with primary
constraints only, while the present work extends the results to the case in
which secondary constraints are also present.

\section{Appendix}

\hspace{\parindent}%
Here we derive the standard covariant Yang-Mills Lagrangian from its
non-covariant version as an example of the procedure of
``covariantizing''
(\ref{lagrang}) to get (\ref{nou lag}). Actually we can directly start
from the Hamiltonian $H_0$ of (\ref{ham}) which for
Yang-Mills takes the form:
$$
H_0 = \frac{1}{2} \pi^k_a\pi^k_a+
\frac{1}{4} F^{ij}_a F^{ij}_a.
$$

We hence identify, in the notation of sect.2
$$
M_{AB} = \delta({\bf x} - {\bf y}) \delta_{ab},\qquad
V(A_i) = \frac{1}{4} F^{ij}_a F^{ij}_a,
$$
and, from equations (\ref{gaugtr2})
\bea
\delta A^a_i = \partial_i \veps^a -
f^{acb}A_i^c \veps^b = {\cal D}_i^{ab}\veps^b \equiv U_{\alpha}^A
\veps^{\alpha},
\nonumber\\
\delta A^a_0 = \partial_0\veps^a + f^{abc}A^c_0 \veps^b
\equiv {\cal D}_0^{ab} \veps^b,
\nonumber
\eea
where we have used the notation
$\lambda^a = A^a_0$ for the Lagrange multipliers, and taken into account
that now $V^{\beta}_{\alpha}(p,q)$, (\ref{ve}), can be chosen to be zero.

Then, we have, for (\ref{nou lag})
$$
L(A_i, \dot{A}_i, A_0) =
\int d^3{x} \,\left[\frac{1}{2}
(\dot{A}_i^a - {\cal D}_i^{ab} A_0^b)
(\dot{A}_i^a - {\cal D}_i^{ab} A_0^b)
- \frac{1}{4} F^{ij}_a F^{ij}_a\right],
$$
and since $(\dot{A}_i^a - {\cal D}_i^{ab} A_0^b)= F^a_{0i}$, we finally
get
$$
\int L\, dx^0 = - \int d^4 x \, \left[
\frac{2}{4} F_{0i}^a F^{0i \, a} +\frac{1}{4} F^{ij}_a F^{ij}_a
\right]=
\int d^4 x \left[-\frac14 F_{\mu \nu} F^{\mu \nu}\right],
$$
that is, the covariant action for Yang-Mills theories.

\section*{Acknowledgements}

The work of J.P., J.M.P. and R.T. has been partially supported by
project no.\,AEN89-0347. The work of C.O. has been supported in part by the
Robert A. Welch Foundation and NSF Grant PHY 9009850.

\endsecteqno

\end{document}